\documentclass[sigconf]{acmart}
\acmConference[Conference'24]{}{July}{2024}
\usepackage{multirow}
\usepackage{graphicx}
\usepackage{subfigure}
\AtBeginDocument{%
  }

\setcopyright{acmlicensed}
\copyrightyear{2024}
\acmYear{2024}
\acmDOI{XXXXXXX.XXXXXXX}

\acmISBN{978-1-4503-XXXX-X/18/06}

\acmSubmissionID{1139}

\begin{document}

\title{System States Forecasting of Microservices with Dynamic Spatio-Temporal Data}

\author{Yifei Xu}
\affiliation{%
  \institution{Key Laboratory of Cyberspace Security Defense, Institute of Information Engineering, CAS, School of Cyber Security, University of Chinese Academy of Sciences}
  \city{Beijing}
  \country{China}
}
\email{xuyifei@iie.ac.cn}

\author{Jingguo Ge}
\authornote{Jingguo Ge and Haina Tang are the corresponding authors.}
\authornotemark[2]
\affiliation{%
  \institution{Key Laboratory of Cyberspace Security Defense, Institute of Information Engineering, CAS, School of Cyber Security, University of Chinese Academy of Sciences}
  \city{Beijing}
  \country{China}}
\email{gejingguo@iie.ac.cn}

\author{Haina Tang}
\authornotemark[2]
\affiliation{%
  \institution{School of Artificial Intelligence, University of Chinese Academy of Sciences}
  \city{Beijing}
  \country{China}}
\email{hntang@ucas.ac.cn}

\author{Shuai Ding}
\affiliation{%
  \institution{Key Laboratory of Cyberspace Security Defense, Institute of Information Engineering, CAS, School of Cyber Security, University of Chinese Academy of Sciences}
  \city{Beijing}
  \country{China}}
\email{dingshuai@iie.ac.cn}

\author{Tong Li}
\affiliation{%
  \institution{Key Laboratory of Cyberspace Security Defense, Institute of Information Engineering, CAS, School of Cyber Security, University of Chinese Academy of Sciences}
  \city{Beijing}
  \country{China}}
\email{litong@iie.ac.cn}

\author{Hui Li}
\affiliation{%
  \institution{Key Laboratory of Cyberspace Security Defense, Institute of Information Engineering, CAS, School of Cyber Security, University of Chinese Academy of Sciences}
  \city{Beijing}
  \country{China}}
\email{lihui1@iie.ac.cn}


\renewcommand{\shortauthors}{Yifei Xu et al.}

\begin{abstract}
In the AIOps (Artificial Intelligence for IT Operations) era, accurately forecasting system states is crucial. In microservices systems, this task encounters the challenge of dynamic and complex spatio-temporal relationships among microservice instances, primarily due to dynamic deployments, diverse call paths, and cascading effects among instances. Current time-series forecasting methods, which focus mainly on intrinsic patterns, are insufficient in environments where spatial relationships are critical. Similarly, spatio-temporal graph approaches often neglect the nature of temporal trend, concentrating mostly on message passing between nodes. Moreover, current research in microservices domain frequently underestimates the importance of network metrics and topological structures in capturing the evolving dynamics of systems. This paper introduces STMformer, a model tailored for forecasting system states in microservices environments, capable of handling multi-node and multivariate time series. Our method leverages dynamic network connection data and topological information to assist in modeling the intricate spatio-temporal relationships within the system. Additionally, we integrate the PatchCrossAttention module to compute the impact of cascading effects globally. We have developed a dataset based on a microservices system and conducted comprehensive experiments with STMformer against leading methods. In both short-term and long-term forecasting tasks, our model consistently achieved a 8.6\% reduction in MAE(Mean Absolute Error) and a 2.2\% reduction in MSE (Mean Squared Error). The source code is available at https://github.com/xuyifeiiie/STMformer.
\end{abstract}




\maketitle

\section{Introduction}
\begin{figure*}[htbp]
\subfigure[]{
\includegraphics[width=1\columnwidth]{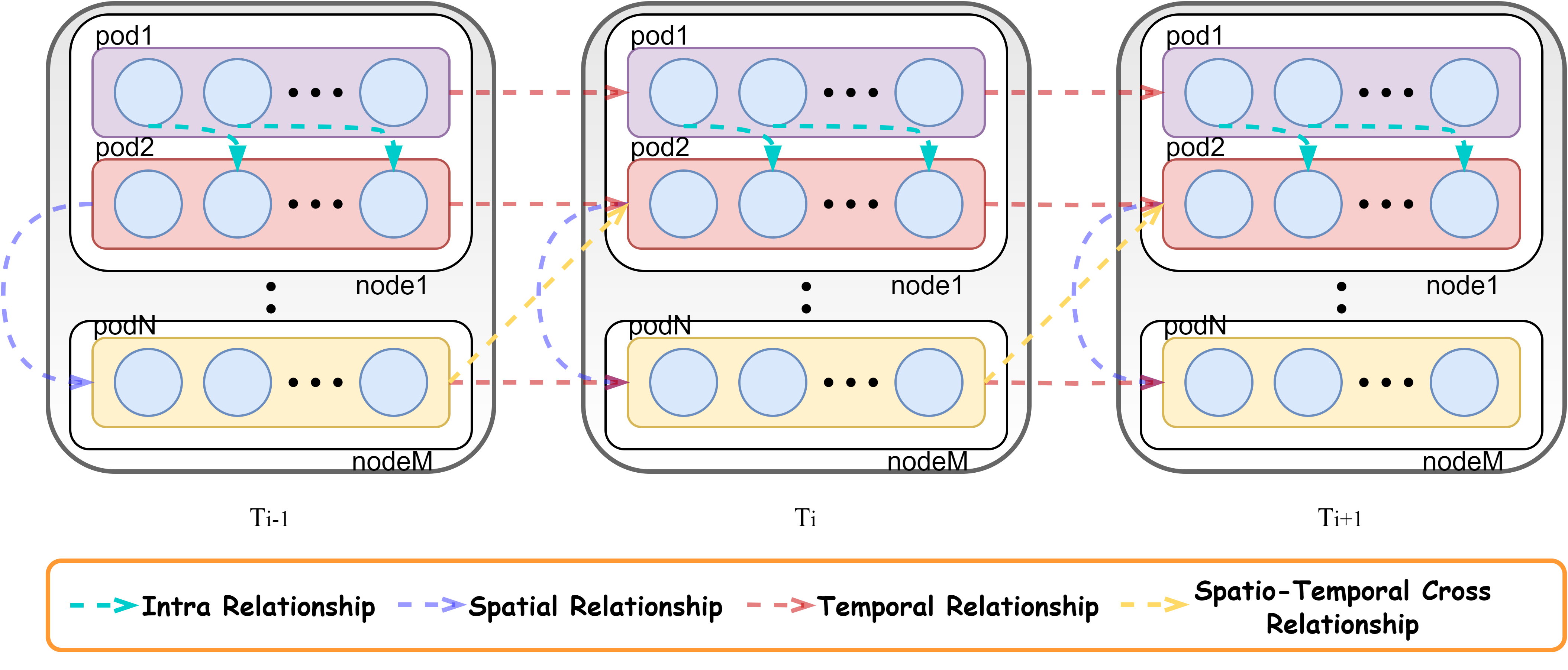}
\label{relationship}}
\subfigure[]{
\includegraphics[width=1\columnwidth]{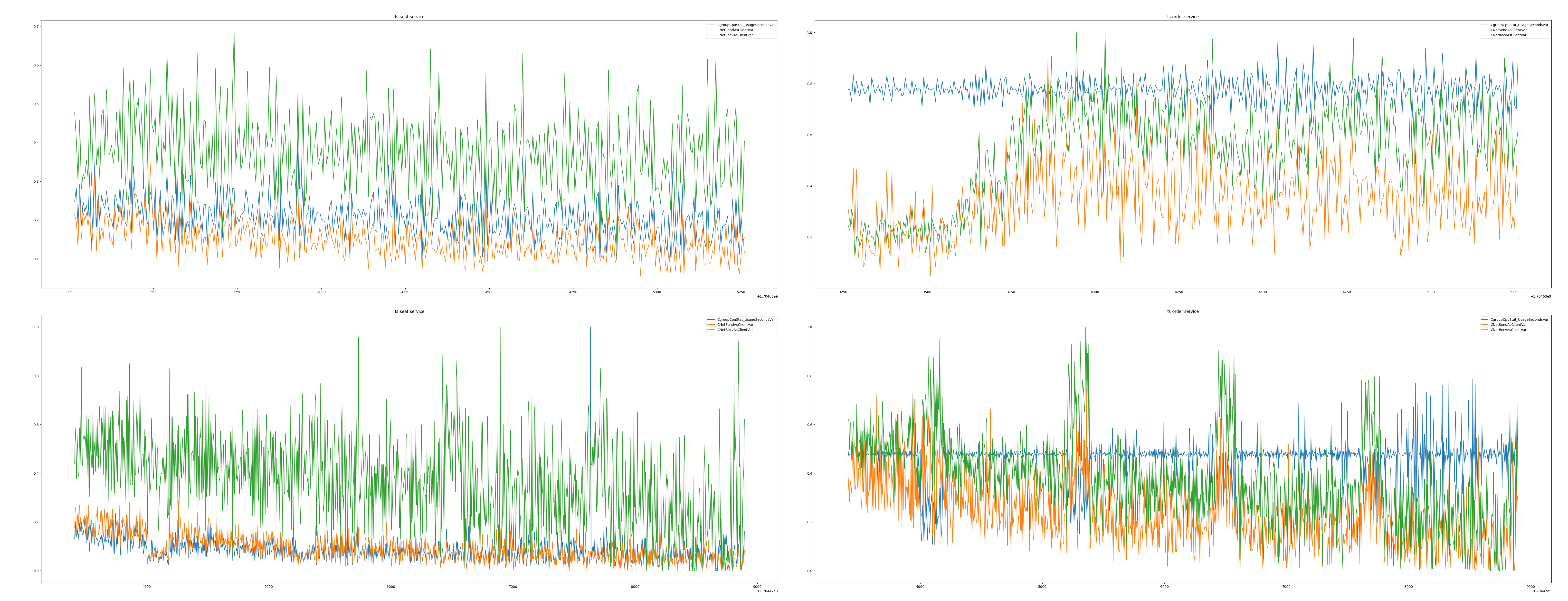}
\label{metrics}}
\caption{(a) shows multiple relationships in the microservices system. The blue circles represents the observation metrics of pod. (b) shows instance metrics of ts-seat-service and ts-order-service under two fault injections. First row shows network loss faults and second row shows CPU stress faults.}
\Description[]{}
\end{figure*}
With the rapid advancement of cloud-native technology, the structural paradigm of cloud applications is transitioning from large monolithic services to distributed microservices systems\cite{balalaie2016migrating}. In production deployments, these microservices could be containerized and deployed over IT infrastructure, greatly expanding the system's vulnerability to both failures and security\cite{hou2021diagnosing, waseem2021nature, pereira2021security}. The escalating complexity and scale of cloud data centers and the growing pressures on SREs(Site Reliability Engineers) are driving the shift towards AIOps(Artificial Intelligence for IT Operations) in industry\cite{mulder2021enterprise}. In the domain of intelligent IT operations, accurate prediction of future system states is essential for downstream tasks, such as anomaly detection and resource allocation, which could grant sufficient time for decision-making and strategy execution\cite{fedushko2020real, ren2023deep}. It epitomizes the sophisticated integration of AI into operational frameworks, emphasizing proactive measures in the dynamic and complex situation of cloud native technologies. 

As illustrated in Fig.~\ref{relationship}, microservice instances are containerized and deployed on nodes (either virtual or physical machines), and these instances are interconnected through network communications that establish dependency relationships, which results in instance states being influenced by neighboring instances in both spatial and temporal dimensions. Fundamentally, system state prediction is a multivariate time series forecasting problem. As research in this field progresses, the current mainstream trend is towards deep learning methods. Informer\cite{zhou2021informer}, Autoformer\cite{wu2021autoformer}, and Fedformer\cite{zhou2022fedformer} explore relationships between time steps from the time or frequency domain and focus on reducing the computational complexity of the attention mechanism. TimesNet\cite{wu2022timesnet} captures sequence features within and between time series cycles. PatchTST\cite{nie2022time} introduces patch operations into time series, enabling the new time-step sequences to contain local information. While these approaches are dedicated to capturing temporal trends, they often lack modeling of the spatial dependencies between instances data. On the other hand, the spatial relationships between instances allow the microservices system to be viewed as a directed graph, enabling predictions using spatio-temporal graph algorithms. STSGCN\cite{song2020spatial} constructs a new adjacency matrix to simultaneously model spatio-temporal relationships within local time steps. Building on this, STSGT\cite{banerjee2022spatial} makes enhancements for global time steps, but it still ignores relationships between nodes from node perspective and the historical time steps of the input sequences are notably short. 


In addition to the issues with existing research methods, the inherent complexity of microservices systems also leads to certain limitations in conducting predictive tasks within this domain. These limitations include: (1) instances on the same node(e.g., pod1 and pod2 in Fig.~\ref{relationship}) can impact each other's performance through resource competition and network congestion\cite{sampaio2019improving,kratzke2015microservices}. Moreover, microservices deployed on platforms such as OpenShift\cite{openshift} and Kubernetes\cite{kubernetes} exhibit dynamism, with service instances frequently scaled, destroyed, and rebuilt based on demand, disrupting prediction outcomes due to changes in deployment structures; (2) the state of microservice instances is influenced not only by the interrelationships among multivariate variables within the instance but also by the performance status of the nodes\cite{liang2021normalized,papana2021connectivity,wang2023dance}. However, existing research\cite{zhang2022deeptralog, gan2019seer, li2018robust, gan2021sage} either employs simplistic network metrics or overlooks node metrics altogether, which are insufficient for aiding in the prediction of critical performance indicators. In the right part of Fig.~\ref{metrics}, the additional network metrics(green lines) exhibited periodic fluctuations following the injection of faults. (3) each access request has a unique call chain, leading to states being influenced by continually changing requests. Static network topology information is inadequate for adapting to the dynamic nature of microservices systems. Moreover, the connection information constructed from trace data lacks the necessary network metrics we require for effective analysis; (4) cascading effects exist within the microservices system, where an anomaly in one instance or a surge in requests can impact upstream and downstream instances. thereby increasing the complexity of spatio-temporal modeling. As the second row of the Fig.~\ref{metrics} shows, CPU usage(blue lines) of ts-order-service occurs delayed fluctuation when we inject CPU stress faults into its adjacent service, ts-seat-service.


Unlike existing studies that use only past observations as input, we believe that additional information can help improve the accuracy of predictions. Therefore, in addition to extra network metrics mentioned above, dynamic topological structure information and adjacency matrices are also essential for predicting system states. Based on OpenTelemetry\cite{opentelemetry}, trace data, which could map the sequence of service or function calls per request, can be collected through distributed tracing tools like Zipkin\cite{zipkin}, Jaeger\cite{jaeger}, and Skywalking\cite{skywalking}, and can be used to generate dynamic connection information. These tools are crucial for visualizing service call chains, providing a foundation for system analysis. However, deploying these solutions faces obstacles. A survey\cite{li2022enjoy} suggests high costs may deter small companies from adoption. Some tools need to invade source code for event tracking or support limited programming languages. Moreover, trace data, which reflects application-level information, fails to capture deeper operational data of instances, pointing to the need for more comprehensive monitoring approaches that combine application interactions with detailed system metrics for a complete view of microservices' performance. To satisfy the aforementioned needs, we have developed a code-free intrusive tool that gathers status information for instances and nodes, along with topological structure information. It utilizes eBPF\cite{ebpf} to trace TCP connections, providing extensive network connection details and dynamic adjacent matrices.

In this study, for system states forecasting tasks, we propose the \textbf{S}patio-\textbf{T}emporal \textbf{M}essage Trans\textbf{former}-based model, STMformer, to address the four previously mentioned challenges: 1) utilizing dynamic topological information to address the dynamic deployment characteristics of microservices, 2) using rich network and node metrics to overcome issues related to insufficient information, 3) employing captured network connection data to construct time-varying adjacency matrices, which assist in accommodating the dynamic changes of microservices instances, and 4) proposing a solution from a global spatio-temporal perspective to address the problem of cascading effect propagation. Our contributions in this paper are listed as follows:
\begin{itemize}
\item We propose a model, STMformer, that utilizes dynamic topological structure information and network connections as auxiliary data. It models the complex spatio-temporal relationships from four perspectives to predict states of the microservices system in both short-term and long-term forecasting tasks.
\item To compute the propagation impacts of cascading effects within the system from a global spatio-temporal perspective, we introduce the PatchCrossAttention module. This module performs attention calculations simultaneously across all time spans of the state sequence and the global topological structure at each time step.
\item Based on an open resource microservices benchmark, we have generated a dataset containing instances and nodes state metrics, dynamic adjacent matrices and topology information. And we evaluate our model against advanced time series methods and spatio-temporal graph methods based on the dataset, achieving superior performance compared to state-of-the-art approaches in short-term and long-term forecasting tasks.
\end{itemize}

This paper is structured as follows. Section 2 gives an overview of the related works. Section 3 details the preliminaries of this paper, and illustrates the structure of our proposed model. Section 4 describes experiments and results, and Section 5 summarizes our conclusions and outlines future work. 

\section{RELATED WORK}

\subsection{Time Series Forecasting}
Time series forecasting emerges as an exceptionally challenging field, requiring a deep understanding of inherent characteristics within temporal trends to predict future outcomes. This discipline demands rigorous analysis and the application of sophisticated models to decode the intricate patterns embedded in time-dependent data, enabling accurate projections of future states. Informer\cite{zhou2021informer} selects top-k weights in attention matrix based on KL-divergence, achieving a substantial reduction in computational complexity and memory usage. Autoformer\cite{wu2021autoformer} presents a novel approach by replacing the conventional self-attention mechanism with an Auto-Correlation function by selecting top-k correlations in correlation matrix. This ingenious substitution exploits the periodic nature of series to identify and aggregate similar sub-series across different periods. Both Informer and Autoformer reduce the computational complexity to $O(L\log{L})$. Fedformer\cite{zhou2022fedformer} employs Fourier and wavelet transformations for enhanced time series analysis, achieving linear complexity and memory efficiency by selectively incorporating crucial frequency components for a holistic representation. TimesNet\cite{wu2022timesnet} proposes a novel approach by identifying the top-k frequencies through Fourier transform and transforming 1D time series data into a 2D space. By leveraging 2D convolutional operations, it captures the temporal variation patterns both within and across different frequency cycles. This methodology achieves optimal performance in a variety of time series tasks, including forecasting, imputation, classification, and anomaly detection, demonstrating its comprehensive applicability and effectiveness in extracting complex temporal dynamics.

\subsection{Spatio-Temporal Graph}
When temporal data also contains spatial relationships, simply analyzing time trends is insufficient. Viewing this kind of data as a spatio-temporal graph has proven beneficial, especially in traffic and weather forecasting, highlighting the importance of considering both spatial and temporal dimensions in predictive analysis.

STAD\cite{yang2022fine} leverage detailed component topology and metric dependencies, integrating dynamic time warping with graph neural networks to capture spatial relationships and temporal dynamics within cloud environments. STSGCN\cite{song2020spatial} innovatively captures localized spatio-temporal correlations and heterogeneities in spatio-temporal data. Unlike previous methods, STSGCN synchronously processes spatial and temporal data through a unified framework, demonstrating superior forecasting performance on real-world datasets. STGIN\cite{luo2022stgin} introduces a spatio-temporal Transformer-based model for long-term traffic forecasting, integrating Graph Attention Network(GAT) and Informer layers to effectively capture the spatial and temporal dependencies of traffic data. However, it relies on static spatial topology for its adjacency matrix and does not incorporate real-time connectivity information into the model. MGT\cite{ye2022meta} innovatively integrates meta-learning with attention mechanisms to model spatio-temporal heterogeneity. But its algorithm heavily relies on for loops, and in the Decoder phase, it adopts a single time-step loop inference, which substantially slows down the model's computational speed. Consequently, spatio-temporal methods more focus on message passing between nodes, neglecting the intrinsic patterns of the time series.

\begin{figure*}[htbp]
\centerline{\includegraphics[width=1\textwidth]{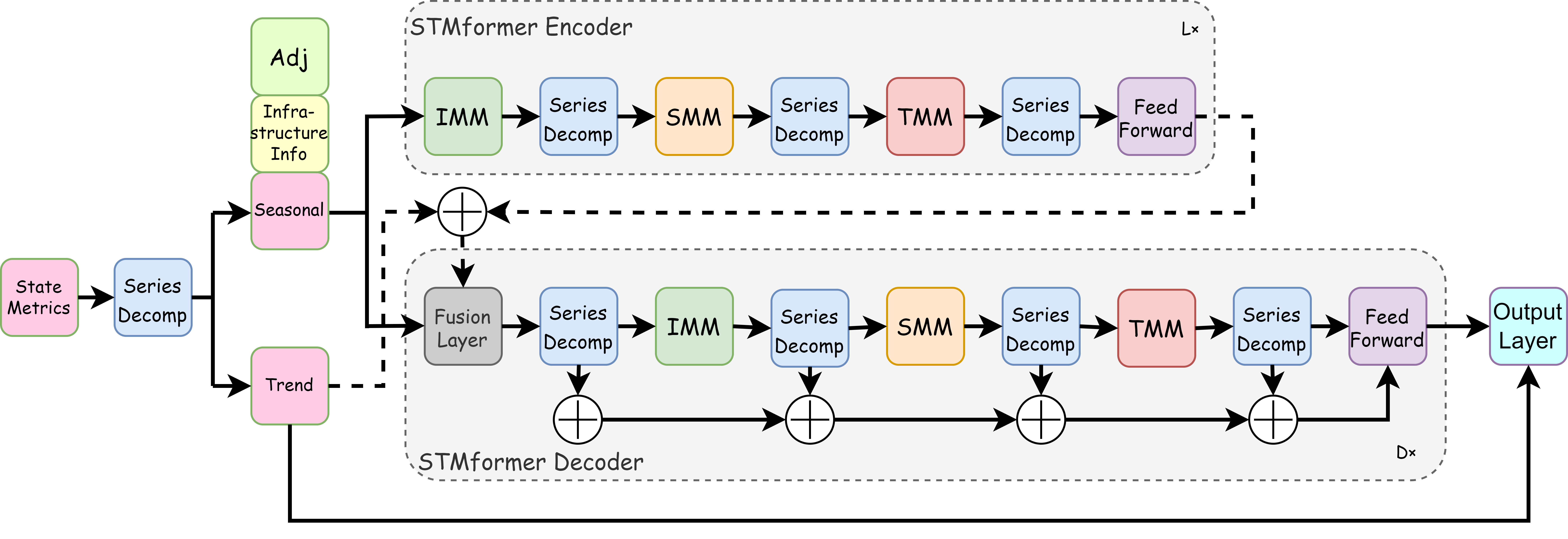}}
\caption{The framework of STMformer. The seasonal component from time series decomposition serves as input to both encoder and decoder. The encoder output is combined with the trend to integrate trend information. In the decoder, this combined output is fused with the original seasonal data to update and refine the state representation.}
\label{model framework}
\Description[]{}
\end{figure*}

\section{METHODOLOGY}
\subsection{Problem Statement}
To elucidate the spatial and temporal dependencies and influences among instances within a microservices system, our framework integrates state metrics with network adjacent matrices and topology information data, enabling us to conceptualize the system architecture as a directed graph. This modeling facilitates a comprehensive analysis of the intricate interplay between various instances, underscoring the critical role of state, network events and topology in understanding the dynamic behavior of microservices. 
\begin{itemize}
\item At the spatial aspect, we represent the microservices system as a directed graph $\mathcal{G}_t=(\mathcal{V}_t,\mathcal{E}_t,\mathcal{A}_t)$ at time step $t$, where $|\mathcal{V}_t|=N$ is the set of nodes representing instances in microservices system; $N$ is the number of nodes and it can change dynamically. For ease of representation, we omit its time subscript and denote it simply as N. $\mathcal{E}_t$ is the set of edges indicating the logic connectivity of each pair of nodes; $\mathcal{A}_t\in{\mathbb{R}^{N\times{N}}}$ denotes the weighted adjacency matrix of the graph at time step $t$. 
\item At the temporal aspect, the nodes' feature matrix of the system can be expressed as $\mathcal{X}^{(t)}_{\mathcal{G}_{t}}\in{\mathbb{R}^{N\times{C}}}$, where $C$ is the dimension of the attribute features of each instance. So the $\mathcal{X}^{(t)}_{\mathcal{G}_t}$ indicates the observed system metrics of graph G at the time step t. 
And $\mathcal{A}^{(t)}_{\mathcal{G}_t}\in{\mathbb{R}^{N\times{N}}}$ is used to denote the connections adjacency matrix at time step $t$, which comes from network connections metrics; $\mathcal{M}^{(t)}_{\mathcal{G}_t}$ denotes the physical position information of pods on hosts.
\item Problem definition: To predict future system state, we need training a function $f$ which maps the history observation metrics $(\mathcal{X}^{(t-T_h+1)}_{\mathcal{G}_{t-T_h+1}},\mathcal{X}^{(t-T_h+2)}_{\mathcal{G}_{t-T_h+2}},...,\mathcal{X}^{(t)}_{\mathcal{G}_{t}})$ with auxiliary data to future system states $(\mathcal{X}^{(t+1)}_{\mathcal{G}_{t+1}},\mathcal{X}^{(t+2)}_{\mathcal{G}_{t+2}},...,\mathcal{X}^{(t+T_p)}_{\mathcal{G}_{t+T_p}})$, where $T_h$ is the length of horizon time steps, $T_p$ is length of prediction time steps. It could be formulized as follows:
\begin{equation}\label{0}
\begin{aligned}
\mathcal{X}_{t+1:t+T_p} &= f(\mathcal{X}_{t-T_h+1:t}, \mathcal{A}_{t-T_h+1:t}, \mathcal{M}_{t-T_h+1:t})
\end{aligned}
\end{equation}
\end{itemize}

\subsection{Model Structure}
The system state forecasting task is to forecast future system state of $T'$ future time steps based on metrics series of $T$ historical time steps. In this section, we will introduce the overall structure of STMformer and its internal module for specific function. The model framework is shown in Fig.~\ref{model framework}. 

\subsubsection{STMformer Framework}
STMformer is a transfomer based model, specifically designed for forecasting system states. As the Fig.~\ref{model framework} shows, STMformer is comprised of an Encoder and a Decoder, each equipped with scalable layers. And IMM, SMM and TMM respectively represents the Intra Message Module, Spatial Message Module and Temporal Message Module. 

\textbf{Model Input of System State Metrics}
Before the state metric series are fed into the Encoder, we implement series decomposition, data embedding and patch operations. Obtaining inspiration from the concept of seasonal-trend decomposition\cite{cleveland1990stl} and single filter is hard to extract trend from series data\cite{zhou2022fedformer}, we dissect the input state metrics into their trend and seasonal components with multi filters. And data embedding operation transforms feature dimension into a higher dimensional vector space. Furthermore, we employ one-stride patch operation to establish connections between information across adjacent time steps, ensuring this process does not alter the time length of the input time series.

Considering there are $N$ pod instances of the microservices, the input system state metrics series of $T$ horizon time steps can be expressed as $\mathcal{X}\in\mathbb{R}^{T\times{N}\times{C}}$, where $C$ is the number of features. 

For the initial series decomposition, we split the state data $\mathcal{X}$ into $\mathcal{X}_s$ and $\mathcal{X}_t$. 
where $\mathcal{X}_{\mathrm{s}},\mathcal{X}_{\mathrm{t}}\in\mathbb{R}^{T\times{N}\times{C}}$ are the seasonal part and trend part respectively. To enable the model to better learn features in the time series beyond trends, we use only $\mathcal{X}_s$ as the input for both the encoder and decoder. We get $\mathcal{X}_{\mathrm{en}}$ and $\mathcal{X}_{\mathrm{de}}\in\mathbb{R}^{T\times{N}\times{D}}$ after applying data embedding and one-stride patch operation on $\mathcal{X}_{\mathrm{s}}$ and $\mathcal{X}_{\mathrm{t}}$. And then we use $\mathcal{X}_\mathrm{it}\in\mathbb{R}^{T\times{N}\times{D}}$ to represent the initial trend part after preprocessing.

\textbf{Model Input of Network Connection Information}
We collected and processed TCP connection events over various time intervals, obtaining key metrics related to the latency of TCP connection establishment and the frequency of these connections between pods or containers. Utilizing these critical data, we constructed adjacency matrices for individual time steps. The adjacency matrices could be denoted as $\mathcal{A}\in\mathbb{R}^{T\times{N}\times{N}}$, and we apply a normalization on $\mathcal{A}$ before input, denoted as $\mathcal{\overline{A}}$.

\textbf{Model Input of Infrastructure Information}
During the collection of system state series, we are concurrently able to acquire instance deployment information. For example, the sequence of system states across node dimensions is sorted by host and pod names. Consequently, pod deployment information is designated as $m_1, m_2, ...$, where $m_i$ represents the count of pods deployed on host $i$. Thus, the input of deployment information could be denoted as $\mathcal{M}\in\mathbb{R}^{T\times{M}\times{1}}$, where $M$ is the number of host.

\textbf{STMformer Structure}
Encoder and Decoder of STMformer are both multi-layer structure. For Encoder, we represent the output of $l$-th encoder layer by $\mathcal{X}^l_{\mathrm{en}} = \text{Encoder}(\mathcal{X}^{l-1}_{\mathrm{en}}, \mathcal{M}, \mathcal{\overline{A}})$, where $l\in\{1,...,L\}$. And $\mathcal{X}^0_{\mathrm{en}} = \mathcal{X}_{\mathrm{en}}$. The $\text{Encoder}(\cdot)$ could be formulized as:

\begin{equation}\label{4}
\begin{aligned}
\mathcal{S}_{\mathrm{en}}^{l, 1},_{-} & =\operatorname{MultiDecomp}\left(\mathrm{IMM}\left(\mathcal{X}_{\mathrm{en}}^{l-1}, \mathcal{M}\right)+\mathcal{X}_{\mathrm{en}}^{l-1}\right) \\
\mathcal{S}_{\mathrm{en}}^{l, 2},_{-} & =\operatorname{MultiDecomp}\left(\mathrm {SMM}\left(\mathcal{S}_{\mathrm{en}}^{l, 1}, \mathcal{\overline{A}}\right)+\mathcal{S}_{\mathrm{en}}^{l, 1}\right) \\
\mathcal{S}_{\mathrm{en}}^{l, 3},_{-} & =\operatorname{MultiDecomp}\left(\mathrm {TMM}\left(\mathcal{S}_{\mathrm{en}}^{l, 2}, \mathcal{\overline{A}}\right)+\mathcal{S}_{\mathrm{en}}^{l, 2}\right)\\
\mathcal{S}_{\mathrm{en}}^{l, 4},_{-} & =\mathrm {FeedForward}\left(\mathcal{S}_{\mathrm{en}}^{l,3}\right)+\mathcal{S}_{\mathrm{en}}^{l, 3}\\
\mathcal{X}_{\mathrm{en}}^{l} & =\mathcal{S}_{\mathrm{en}}^{l, 4}
\end{aligned}
\end{equation}
where $\mathcal{S}_{\mathrm{en}}^{l, i}\in\mathbb{R}^{T\times{N}\times{D}},i\in\{1,2,3,4\}$ denotes the seasonal component after the $i$-th series decomposition block in the $l$-th layer. The Infrastructure Message Module(IMM), Spatial Message Module(SMM) and Temporal Message Module(TMM) will be discussed in the following sections.


Like Encoder, the output of $l$-th decoder layer could be represented as $\mathcal{X}^l_{\mathrm{de}} = \text{Decoder}(\mathcal{X}^{l-1}_{\mathrm{de}}, \mathcal{X}_{\mathrm{en}}^{L}, \mathcal{M}, \mathcal{\overline{A}})$, where $l\in\{1,...,L'\}$. $\mathcal{X}^0_{\mathrm{de}} = \mathcal{X}_{\mathrm{de}} + \mathcal{X}_\mathrm{it}$. The $\text{Decoder}(\cdot)$ could be formulized as:

\begin{equation}\label{5}
\begin{aligned}
\mathcal{S}_{\mathrm{de}}^{l, 1},\mathcal{T}_{\mathrm{de}}^{l, 1} & =\operatorname{MultiDecomp}\left(\mathrm{Attention}\left(\mathcal{X}_{\mathrm{en}}^{L}, \mathcal{X}_{\mathrm{de}}^{l-1}\right)+\mathcal{X}_{\mathrm{de}}^{l-1}\right) \\
\mathcal{S}_{\mathrm{de}}^{l, 2},\mathcal{T}_{\mathrm{de}}^{l, 2} & =\operatorname{MultiDecomp}\left(\mathrm{IMM}\left(\mathcal{S}_{\mathrm{de}}^{l, 1}, \mathcal{M}\right)+\mathcal{S}_{\mathrm{de}}^{l, 1}\right) \\
\mathcal{S}_{\mathrm{de}}^{l, 3},\mathcal{T}_{\mathrm{de}}^{l, 3} & =\operatorname{MultiDecomp}\left(\mathrm {SMM}\left(\mathcal{S}_{\mathrm{de}}^{l, 2}, \mathcal{\overline{A}}\right)+\mathcal{S}_{\mathrm{de}}^{l, 2}\right) \\
\mathcal{S}_{\mathrm{de}}^{l, 4},\mathcal{T}_{\mathrm{de}}^{l, 4} & =\operatorname{MultiDecomp}\left(\mathrm {TMM}\left(\mathcal{S}_{\mathrm{de}}^{l, 3}, \mathcal{\overline{A}}\right)+\mathcal{S}_{\mathrm{de}}^{l, 3}\right)\\
\mathcal{S}_{\mathrm{de}}^{l, 5} & =\mathrm {FeedForward}\left(\mathcal{S}_{\mathrm{de}}^{l, 4}\right)+\mathcal{S}_{\mathrm{de}}^{l, 4}\\
\mathcal{T}_{\mathrm{de}}^{l} & = \mathcal{W}_{l,1}\cdot\mathcal{T}_{\mathrm{de}}^{l, 1}+\mathcal{W}_{l,2}\cdot\mathcal{T}_{\mathrm{de}}^{l, 2}+\mathcal{W}_{l,3}\cdot\mathcal{T}_{\mathrm{de}}^{l, 3}+\mathcal{W}_{l,4}\cdot\mathcal{T}_{\mathrm{de}}^{l, 4}\\
\mathcal{X}_{\mathrm{de}}^{l} & =\mathcal{S}_{\mathrm{de}}^{l, 5} + \mathcal{T}_{\mathrm{de}}^{l}
\end{aligned}
\end{equation}


where $\mathcal{S}_{\mathrm{de}}^{l, i},\mathcal{T}_{\mathrm{de}}^{l, j}\in\mathbb{R}^{T\times{N}\times{D}},i\in\{1,2,3,4,5\},j\in\{1,2,3,4\}$ denotes the seasonal and trend component after the $i$-th series decomposition block in the $l$-th layer respectively. $\mathcal{W}_{l,i},i\in\{1,2,3,4\}$ denotes the projector for the $i$-th extracted trend $\mathcal{T}_{\mathrm{de}}^{l, i}$.
The output layer of STMformer aggregate seasonal and trend parts as $\mathcal{W}_{o}\cdot(\mathcal{X}_{\mathrm{de}}^{L'}+\mathcal{X}_\mathrm{it})$, where $\mathcal{W}_{o}$ is the projector to map the $(\mathcal{X}_{\mathrm{de}}^{L'}+\mathcal{X}_\mathrm{it})$ from dimension $D$ to original feature dimension $C$. Attention module here we use is ProbSparseAttention\cite{zhou2021informer}.

\subsubsection{Infrastructure Message Module(IMM)}
While instances deployed on the same virtual machine are isolated in terms of resources through cgroups, there may exists a notable impact on network performance among these pods. This module focuses on applying an attention mechanism to learn from the interactions between pods situated on the same host. And IMM module is also multi-layer-based, and $\mathcal{I}^{\mathrm{0}} = \mathcal{X}_{\mathrm{en}}^{l-1}$ or $\mathcal{S}_{\mathrm{de}}^{l, 1}$. Its computation process could be formulized as 
\begin{equation}\label{6}
\begin{aligned}
\mathcal{I}_\mathrm{1}, \mathcal{I}_\mathrm{2}, ....,\mathcal{I}_\mathrm{M} & = \operatorname{Split}\left(\mathcal{I}^{\mathrm{j-1}},\mathcal{M}\right)\\
\text{Attn}_\mathrm{m} & = \text{Attention}\left(\mathcal{I}_\mathrm{m}\right) \\
\mathcal{I}^{\mathrm{j}} & = \operatorname{Concat}({\text{Attn}_{1},\text{Attn}_{2},...\text{Attn}_\mathrm{m}})
\end{aligned}
\end{equation}
where $\mathcal{I}^\mathrm{j},j\in{1,2,...,K_{1}}$ denotes output of the $j$-th layer of IMM, $K_{1}$ is the number of layers for IMM, and $\mathcal{I}_\mathrm{m}$ is the $m$-th split result.

\subsubsection{Spatial Message Module(SMM)}
The interaction between instances through business request calls establishes a logical topology, delineating a network of dependencies and communications. Within a given time step, the spatial relationships among instances, influenced by network interactions within that timeframe, become pivotal. To capture and dynamically update the topological structure and inter-node relationships based on these interactions, we employ Graph Attention Networks (GAT). And SMM module is also multi-layer-based, and $\mathcal{S}^{\mathrm{0}} = \mathcal{S}_{\mathrm{en}}^{l,1}$ or $\mathcal{S}_{\mathrm{de}}^{l, 2}$. The process is formulized as
\begin{equation}\label{7}
\begin{aligned}
\mathcal{S}^{\mathrm{j}} & = \operatorname{GAT}\left(\mathcal{S}^{\mathrm{j-1}},\mathcal{\overline{A}}\right)
\end{aligned}
\end{equation}
where $\mathcal{S}^\mathrm{j}\in\mathbb{R}^{T\times{N}\times{D}},j\in{1,2,...,K_{2}}$ denotes output of the $j$-th layer of SMM, $K_{2}$ is the number of layers for SMM.

\subsubsection{Temporal Message Module(TMM)}
Time series inherently exhibit trend variations based on historical data. However, when these time series are interrelated spatially, the spatial influence among the observed metric series data within instances also necessitates consideration. In light of this, our model structure incorporates two distinct modules: TB(TimesBlock) to capture the intrinsic trend changes of the time series and PCA(PatchCrossAttention) to account for the alterations induced by spatial influences on the time series. And $\mathcal{T}^{\mathrm{0}} = \mathcal{S}_{\mathrm{en}}^{l,2}$ or $\mathcal{S}_{\mathrm{de}}^{l, 3}$.
\begin{equation}\label{8}
\begin{aligned}
\mathcal{T}^{\mathrm{j}} & = \alpha\cdot\operatorname{TB}\left(\mathcal{T}^{\mathrm{j-1}},\mathcal{\overline{A}}\right)+(1-\alpha)\cdot\operatorname{PCA}\left(\mathcal{T}^{\mathrm{j-1}},\mathcal{\overline{A}}\right)
\end{aligned}
\end{equation}
where $\mathcal{T}^\mathrm{j}\in\mathbb{R}^{T\times{N}\times{D}},j\in{1,2,...,K_{3}}$ denotes output of the $j$-th layer of TMM, $K_{3}$ is the number of layers for TMM, and $\alpha\in\mathbb{R}^{T\times{N}\times{N}}$ is a uniform random matrix. TB and PCA module are both multi-layer-based.

\textbf{TimesBlock(TB)}
To capture the intrinsic variations within time series, our model employs the state-of-the-art TimesBlock\cite{wu2022timesnet} module. This module leverages the Fast Fourier Transform(FFT) to identify the top-k frequencies with the highest significance. Subsequent convolution operations are applied to extract both intra-period and inter-period information from these selected frequencies. This methodology significantly enhances our ability to discern the underlying trends and patterns within the time series data. The computation process could be formulized as
\begin{equation}\label{9}
\begin{aligned}
\mathcal{ST}^{\mathrm{j}} & = \operatorname{TB}\left(\mathcal{ST}^{\mathrm{j-1}}_{\mathrm{TB}},\mathcal{\overline{A}}\right)
\end{aligned}
\end{equation}
where $\mathcal{ST}^{\mathrm{j}}_{\mathrm{TB}}\in\mathbb{R}^{T\times{N}\times{D}},j\in{1,2,...,K_3}$ denotes output of the $j$-th layer of TB.
To facilitate the computation, we get weights of nodes by summing in-degree and out-degree of each node base on $\mathcal{\overline{A}}$, which will eliminate the dimension $N$. 

\begin{figure*}[htbp]
\centerline{\includegraphics[width=1\textwidth]{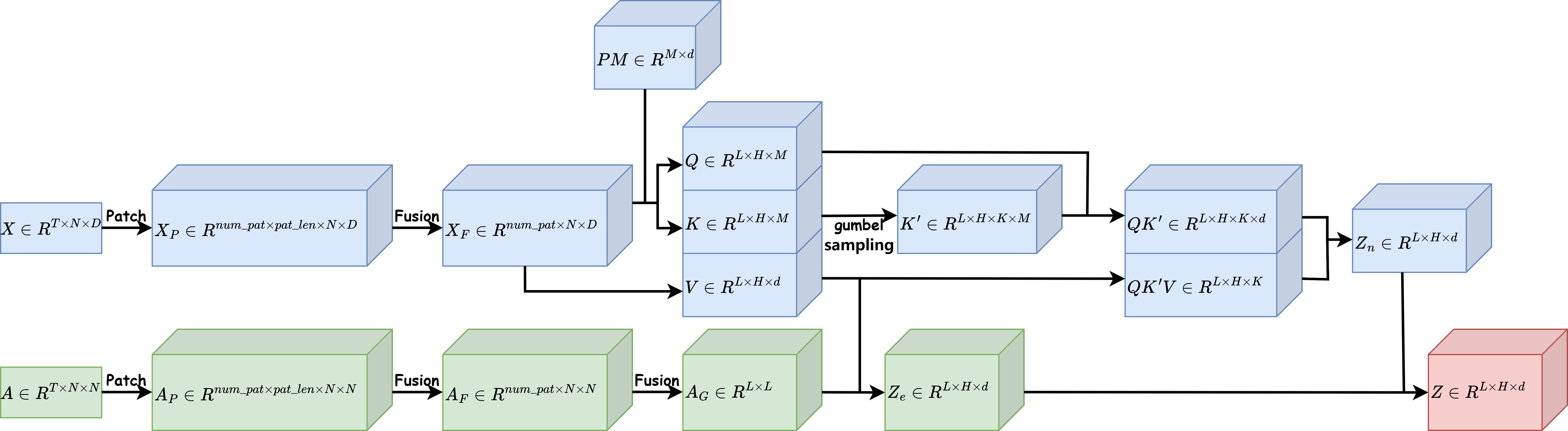}}
\caption{The procedure PatchCrossAttention of to compute global attention among all time steps and nodes. The upper chain in blue is to compute global attention by nodes. The lower chain in green is to compute global attention by edges.}
\label{pcah}
\Description[]{}
\end{figure*}

\textbf{PatchCrossAttention(PCA)}
In the intricate landscape of microservice architectures, a fault or surge in one service on the call chain does not immediately impact its upstream or downstream counterparts. Instead, there exists a delay in the propagation of such faults, leading to a cascading effect. Additionally, adjacent nodes across multiple hops may be affected after a more extended time interval. To address this issue, we consider solutions from two perspectives: one from the node perspective and the other from the edge perspective. From the node perspective, it is crucial for a single node to engage in attention mechanism computations with all nodes across multiple time steps. However, this requirement can lead to substantial computational overhead and may not effectively correlate trends across time steps. From the edge perspective, we consider constructing a global spatio-temporal adjacency matrix to refine the computational attention results of the nodes.

To overcome these challenges, we introduce the PatchCrossAttention Module. The computation process of PCA is shown in Fig.~\ref{pcah}. The upper blue section represents global spatio-temporal attention computations at the node level, while the lower green section involves corrections to the node-level computational results based on the global spatio-temporal adjacency matrix.


In the node-level perspective, despite implementing patch operations to reduce the temporal dimension from $T$ to $P$, where $P<T$, the computational burden of calculating attention across nodes for different time segments still remains. Inspired by Nodeformer\cite{wu2022nodeformer}, we employ the approach that integrates kernel function approximation mapping alongside Gumbel-Softmax for sampling. This approach facilitates the reduction of attention computation to a linear complexity. The process of PCA could be fomulized as follows:
\begin{equation}\label{10}
\begin{aligned}
\mathcal{ST}^{\mathrm{j}}_{\mathrm{PCA}} & = \operatorname{PCA}\left(\mathcal{ST}^{\mathrm{j-1}}_{\mathrm{PCA}},\mathcal{\overline{A}}\right)
\end{aligned}
\end{equation}
where $\mathcal{ST}^{\mathrm{j}}_{\mathrm{PCA}}\in\mathbb{R}^{T\times{N}\times{D}},j\in{1,2,...,K_3}$ denotes output of the $j$-th layer of PCA.

Concretely, using the Positive Random Features(PRF)\cite{krzysztof2021rethinking} to transform original $Q, K, V$ matrix into a low-dimensional vector space. Then the attention formula could be represented as:
\begin{equation}\label{11}
\begin{aligned}
\phi(\mathbf{x})& =\frac{\exp \left(\frac{-\|\mathbf{x}\|_{2}^{2}}{2}\right)}{\sqrt{m}}\left[\exp \left(\mathbf{w}_{1}^{\top} \mathbf{x}\right), \cdots, \exp \left(\mathbf{w}_{m}^{\top} \mathbf{x}\right)\right] \\
\mathbf{z}_{u} & =\sum_{v=1}^{N} \frac{\phi\left(\mathbf{q}_{u}\right)^{\top} \phi\left(\mathbf{k}_{v}\right)}{\sum_{w=1}^{N} \phi\left(\mathbf{q}_{u}\right)^{\top} \phi\left(\mathbf{k}_{w}\right)} \cdot \mathbf{v}_{v} \\ & =\frac{\sum_{v=1}^{N} \phi\left(\mathbf{k}_{v}\right) \cdot \mathbf{v}_{v}^{\top}}{\sum_{w=1}^{N} \phi\left(\mathbf{k}_{w}\right)}
\end{aligned}
\end{equation}
where $\mathbf{z}_{u}$ is the attention weight from node $u$ to node $v$, $\mathbf{w}_{k} \sim \mathcal{N}\left(0, I_{d}\right)$ is independent-identically-distributed, sampled random transformation. $m$ is the dimension of the mapping vector space. 

For mitigating the over-normalization problem and avoid discontinuity of argmax function, using reparametrization trick 
with Gumbel distribution for 
approximation, \eqref{11} will be rewrite as:
\begin{equation}\label{12}
\begin{aligned}
\mathbf{z}_{u} & \approx \sum_{v=1}^{N} \frac{\phi\left(\mathbf{q}_{u} / \sqrt{\tau}\right)^{\top} \phi\left(\mathbf{k}_{v} / \sqrt{\tau}\right) e^{g_{v} / \tau}}{\sum_{w=1}^{N} \phi\left(\mathbf{q}_{u} / \sqrt{\tau}\right)^{\top} \phi\left(\mathbf{k}_{w} / \sqrt{\tau}\right) e^{g_{w} / \tau}} \cdot \mathbf{v}_{v} \\ & =\frac{\sum_{v=1}^{N} e^{g_{v} / \tau} \phi\left(\mathbf{k}_{v} / \sqrt{\tau}\right) \cdot \mathbf{v}_{v}^{\top}}{\sum_{w=1}^{N} e^{g_{w} / \tau} \phi\left(\mathbf{k}_{w} / \sqrt{\tau}\right)}
\end{aligned}
\end{equation}
where $g_{u}$ is independent-identically-distributed sampled from Gumbel distribution and $\tau$ is a temperature coefficient. And this sampling operation could be repeated for $k$ times, which is a hyperparameter infecting the prediction performance and would be discussed in Section 4.


After above computation, we would get new system state $\mathcal{Z}^{\mathrm{j}}\in\mathbb{R}^{PN\times{D}}$. Then we construct a global spatio-temporal adjacency matrix as STSGT\cite{banerjee2022spatial} did for edge-level refinement, denoted as $\mathcal{A}_\mathrm{ST}\in\mathbb{R}^{PN\times{PN}}$, which is used for refine the attention result by connection message. And $\mathcal{A}_\mathrm{ST}$ has global scope by incorporating all time steps within a batch. The diagonal matrices within this global matrix are the result of aggregating attention calculations across multiple time steps within a single patch, yielding a synthesized adjacency matrix. Adjacent to the diagonal, identity matrices are utilized to represent the logical connections of adjacent time segments to themselves, thereby acknowledging the inherent connectivity within and across different time frames. 
This process is fomulized as follows:
\begin{equation}\label{13}
\begin{aligned}
\mathcal{Z}^{\mathrm{j}} & = \operatorname{LayerNorm}\left(\mathcal{A}_\mathrm{ST}\right) \cdot \mathcal{Z}^{\mathrm{j}} \\
\mathcal{ST}^{\mathrm{j}}_{\mathrm{PCA}} & = \operatorname{Flatten}\left(\mathcal{Z}^{\mathrm{j}}\right)
\end{aligned}
\end{equation}

\section{Experiment}
In this section, we will discuss data collection and processing, provide an overview of the benchmark model, and analyze the results of comparative experiments, ablation studies, and hyperparameter studies.

\subsection{Dataset}
Based on the coroot-node-agent\cite{coroot}, we developed an distributed metrics collector to obtain the data we need. The collector could not only integrate eBPF programs, enabling the tracking of TCP connection events among instances, including but not limited to connection establishment, listening, drop and retransmission activities, but also record metrics of pods and nodes with cgroup technique. We define more than 80 kinds of metrics, including CPU usage, memory usage, disk I/O stats, connection latency, span of TCP states transferring, length of socket queue and throughput of sockets and so on.

\begin{table*}[htbp]
\caption{Performance comparison of different baselines based on system states dataset for short-term forecasting tasks. Bold scores and red scores indicate the best and the second best, respectively. Lower metrics indicates better performance.}
\label{short}
\resizebox{\textwidth}{!}{
\begin{tabular}{|cc|c|c|c|c|c|c|c|c|c|}
\hline
\multicolumn{2}{|c|}{Baseline methods} &  &  &  &  &  &  &  &  &  \\ \cline{1-2}
\multicolumn{1}{|c|}{Time Steps} & Metric & \multirow{-2}{*}{Informer} & \multirow{-2}{*}{Autoformer} & \multirow{-2}{*}{FEDformer} & \multirow{-2}{*}{PatchTST} & \multirow{-2}{*}{TimesNet} & \multirow{-2}{*}{DLinear} & \multirow{-2}{*}{STSGCN} & \multirow{-2}{*}{STSGT} & \multirow{-2}{*}{STMformer} \\ \hline
\multicolumn{1}{|c|}{} & MAE & 0.04639 & 0.04794 & 0.01709 & {\color[HTML]{FE0000} 0.01915} & 0.01932 & 0.03786 & 0.05113 & 0.6793 & {\color[HTML]{000000} \textbf{0.01663}} \\ \cline{2-2}
\multicolumn{1}{|c|}{} & MSE & 0.01441 & 0.006213 & 0.002774 & {\color[HTML]{FE0000} 0.002126} & 0.002993 & 0.004717 & 0.0163 & 283.221 & \textbf{0.002102} \\ \cline{2-2}
\multicolumn{1}{|c|}{\multirow{-3}{*}{16}} & RMSE & 0.12011 & 0.07841 & 0.05202 & {\color[HTML]{FE0000} 0.04577} & 0.05408 & 0.06858 & 0.1276 & 5.062 & \textbf{0.04537} \\ \hline
\multicolumn{1}{|c|}{} & MAE & 0.04702 & 0.05844 & 0.0177 & 0.01912 & {\color[HTML]{FE0000} 0.01684} & 0.03208 & 0.05158 & 0.4264 & \textbf{0.01623} \\ \cline{2-2}
\multicolumn{1}{|c|}{} & MSE & 0.01443 & 0.008268 & 0.0027 & {\color[HTML]{FE0000} 0.002135} & 0.002549 & 0.004362 & 0.01624 & 0.2968 & {\color[HTML]{000000} \textbf{0.002084}} \\ \cline{2-2}
\multicolumn{1}{|c|}{\multirow{-3}{*}{32}} & RMSE & 0.12034 & 0.09054 & 0.0514 & {\color[HTML]{FE0000} 0.04587} & 0.04976 & 0.06593 & 0.1274 & 0.5448 & {\color[HTML]{000000} \textbf{0.04516}} \\ \hline
\end{tabular}}
\end{table*}

Besides, we collect data from a microservices benchmark, Train Ticket\cite{ticket}. It has 41 microservices and programmed by various program languages. For workload generator, we write a script with locust\cite{locust} to simulate behaviors of users visiting different microservices. To emulate microservices system performance under varied loads, we adjusted concurrent user numbers and utilized chaos-mesh\cite{chaos-mesh}, a cloud-native fault injection platform, to introduce six specific faults. This process enabled the collection of detailed system metrics in different fault conditions and normal status, offering insights into the system's resilience and behavior under stress. We record metrics every 5 seconds and spent 1 to 2 hours under different conditions. Then we sort and aggregate data from nodes by time stamp. Taking long-term forecasting as example, the dataset covers about 14000 samples, every sample is an array consisting 64 time steps, 56 pods and 80 features.

Our deployed benchmark is an open-source experimental system whose components have not undergone production-level testing. As a result, under high concurrency and fault injection conditions, system stability cannot be guaranteed, potentially leading to extreme values in network observation data. To speed up computation convergence, we apply log1p to network-relevant metrics and use min-max normalization to scale data to $[0,1]$, both of which are reversible processes.

\subsection{Baselines}
We compare our proposed STMformer against eight baseline models on short-term and long-term system states forecasting tasks. The baseline models can be divided into two main categories depending on how they process time series. The first category integrates the metrics of all instances, ignoring the spatial relationship, which means the input size is $X\in\mathbb{R}^{T\times{NC}}$. It includes Informer\cite{zhou2021informer}, Autoformer\cite{wu2021autoformer}, FEDformer\cite{zhou2022fedformer}, PatchTST\cite{nie2022time}, TimesNet\cite{wu2022timesnet}, and DLinear\cite{zeng2023transformers}. The second category are spatio-temporal methods, considering temporal and spatial relationship between instances, such as STSGCN\cite{song2020spatial} and STSGT\cite{banerjee2022spatial}. Its input size is $X\in\mathbb{R}^{T\times{N}\times{C}}$. 

\begin{itemize}
    \item Informer: A Transformer-based model using ProbSparseAttention and distillation to accelerate computation for time series prediction.
    \item Autoformer: A Transformer-based model using series decomposition and auto correlation to capture pattern of time trend.
    \item FEDformer: A Transformer-based model selecting components randomly to represent time series and calculating attention in frequency domain.
    \item PatchTST: A Transformer-based model using patching to aggregate local information from time series and perform channel-independence computation.
    \item TimesNet: A CNN-based model leveraging FFT to find top-k the most relevant frequencies and applying CNN on time series by capturing time pattern within or across various cycles.
    \item DLinear: A Linear model for time series prediction.
    \item STSGCN: A GCN-based model by using localized spatio-temporal subgraph module to find localized correlations independently. 
    \item STSGT: A model uses GCN to apply spatial message passing and calculates attention by building a spatial-temporal synchronous graph among all nodes and time steps.
\end{itemize}

\begin{table*}[htbp]
\caption{Performance comparison of different baselines based on system states dataset for long-term forecasting tasks. Bold scores and red scores indicate the best and the second best, respectively. Lower metrics indicates better performance.}
\label{long}
\resizebox{\textwidth}{!}{
\begin{tabular}{|cc|c|c|c|c|c|c|c|c|c|}
\hline
\multicolumn{2}{|c|}{Baseline methods} &  &  &  &  &  &  &  &  &  \\ \cline{1-2}
\multicolumn{1}{|c|}{Time Steps} & Metric & \multirow{-2}{*}{Informer} & \multirow{-2}{*}{Autoformer} & \multirow{-2}{*}{FEDformer} & \multirow{-2}{*}{PatchTST} & \multirow{-2}{*}{TimesNet} & \multirow{-2}{*}{DLinear} & \multirow{-2}{*}{STSGCN} & \multirow{-2}{*}{STSGT} & \multirow{-2}{*}{STMformer} \\ \hline
\multicolumn{1}{|c|}{} & MAE & 0.2062 & 0.03179 & 0.03156 & {\color[HTML]{FE0000} 0.01899} & 0.02914 & 0.04732 & 0.05673 & 0.07784 & \textbf{0.01644} \\ \cline{2-2}
\multicolumn{1}{|c|}{} & MSE & 0.07041 & 0.004087 & 0.003949 & {\color[HTML]{FE0000} 0.002053} & 0.003774 & 0.006743 & 0.01656 & 63.707 & \textbf{0.002039} \\ \cline{2-2}
\multicolumn{1}{|c|}{\multirow{-3}{*}{64}} & RMSE & 0.2653 & 0.06382 & 0.06273 & {\color[HTML]{FE0000} 0.04523} & 0.06132 & 0.07915 & 0.1287 & 6.275 & \textbf{0.04460} \\ \hline
\multicolumn{1}{|c|}{} & MAE & 0.04898 & 0.08693 & 0.02627 & 0.0184 & {\color[HTML]{FE0000} 0.01818} & 0.02241 & - & - & {\color[HTML]{330001} \textbf{0.01728}} \\ \cline{2-2}
\multicolumn{1}{|c|}{} & MSE & 0.01450 & 0.01618 & 0.002893 & \textbf{0.002081} & 0.002239 & 0.002435 & - & - & {\color[HTML]{FE0000} 0.002126} \\ \cline{2-2}
\multicolumn{1}{|c|}{\multirow{-3}{*}{128}} & RMSE & 0.1204 & 0.1264 & 0.05329 & \textbf{0.04521} & 0.04663 & 0.04779 & - & - & {\color[HTML]{FE0000} 0.04642} \\ \hline
\end{tabular}}
\end{table*}

\subsection{Experiment Setups}
Train Ticket microservices are deployed on seven virtual machines, each of which has a 8 GB RAM with system version Ubuntu 20.04.5 LTS. Each model is trained on a single NVIDIA A100 80G GPUs.

Dataset of all categories are divided into training sets, validation sets, and test sets in a 8:1:1 ratio during modeling. We set the horizon steps same as prediction steps. And we conduct short-term and long-term prediction experiments on step size $L_{short} = [16, 32]$ and $L_{long} = [64, 128]$ respectively. All transformer-based model are set up to contain 3 encoder layers and 1 decoder layer. For detail of training, model is refined by MSE loss and ADAM optimizer. We also use gradient accumulation and warming up strategy for small batches training, since each sample is a 3-dimension array. 

\subsection{Overall Performance}
To comprehensively evaluate the performance of our model in system state prediction tasks, we conducted experiments for both short-term and long-term forecasting and compared them with benchmark models. Table \ref{short} displays the results of the short-term forecasting tasks, while Table \ref{long} shows the results of the long-term forecasting tasks.

Based on the experimental results, we can draw the following conclusions:
(1) In short-term forecasting tasks, our model outperforms other benchmark models. In long-term tasks, our model slightly surpasses PatchTST in some metrics while generally performing better than the benchmark models. Our model improves upon the next best model by 8.6\% in MAE and 2.2\% in MSE. (2) Time series models demonstrate superior performance in capturing temporal trends, but they lack modeling of spatial relationships, which significantly impacts system state prediction tasks for microservices. The results from TimesNet exceed those of our model, and our use of the TimesBlock module in our model indicates that the inability to transfer spatial information between nodes affects the task outcomes. (3) Spatio-temporal graph models generally show poor performance, especially the STSGT model, which exhibits non-convergence issues at strides of 16 and 64. The main reason could be that it does not consider the correlation of node states at different time steps from the perspective of the nodes themselves, but only performs message passing from the perspective of the edges. Meanwhile, STSGCN shows performance similar to Informer but fails to capture intrinsic patterns of the time series, hence its performance is lower than our model. Finally, STMformer can accurately capture dynamic features in spatio-temporal data and effectively compensates for the shortcomings of these methods, thereby achieving superior performance.


\subsection{Ablation Studies}
STMformer contains four key components:IMM, SMM, TMM-TB, TMM-PCA. To prove that the key modules can help the STMformer to mine spatio-temporal feature of data, we designed five variants of the model. These variants were compared against the whole STMformer, with sharing identical hyperparameters. The descriptions of these five variants are as follows:
\begin{table}[htbp]
\caption{Ablation experiments of STMformer based on system states dataset long-term forecasting tasks. Bold scores indicate the best. Lower metrics indicates better performance.}
\label{ablation}
\begin{tabular}{|cc|ccc|}
\hline
\multicolumn{1}{|c|}{\multirow{2}{*}{Methods}} & Time Steps & \multicolumn{3}{c|}{64}                                    \\ \cline{2-5} 
\multicolumn{1}{|c|}{}                         & Metric     & \multicolumn{1}{c|}{MAE} & \multicolumn{1}{c|}{MSE} & RMSE \\ \hline
\multicolumn{2}{|c|}{w/o IMM}                               & \multicolumn{1}{c|}{0.5768} & \multicolumn{1}{c|}{0.01691} & 0.1300  \\ \hline
\multicolumn{2}{|c|}{w/o SMM}                               & \multicolumn{1}{c|}{0.1625} & \multicolumn{1}{c|}{0.06997} & 0.2644  \\ \hline
\multicolumn{2}{|c|}{w/o TMM-TB}                            & \multicolumn{1}{c|}{0.06339}  & \multicolumn{1}{c|}{0.02082}  & 0.1443   \\ \hline
\multicolumn{2}{|c|}{w/o TMM-PCA}                           & \multicolumn{1}{c|}{0.04434}  & \multicolumn{1}{c|}{0.01183}  & 0.1087   \\ \hline
\multicolumn{2}{|c|}{w/o Adjacency}                         & \multicolumn{1}{c|}{0.09318}  & \multicolumn{1}{c|}{0.02471}  & 0.1571   \\ \hline
\multicolumn{2}{|c|}{STMformer}                             & \multicolumn{1}{c|}{\textbf{0.01644}}  & \multicolumn{1}{c|}{\textbf{0.002039}}  & \textbf{0.0446}   \\ \hline
\end{tabular}
\end{table}

\begin{itemize}
\item w/o IMM: This model variant removes the IMM. It is used to assess the influence of metrics among microservice instances deployed on the same virtual node.

\item w/o SMM: This model variant removes the SMM. It is utilized to evaluate the impact of dynamic connection information between instances at a single time step on the state of the instances.

\item w/o TMM-TB: This model variant removes the TB module of TMM, serving to demonstrate the necessity of learning the trends in its own time series variations.

\item w/o TMM-PCA: This model variant eliminates the PCA module from the TMM, aiming to illustrate the critical importance of conducting a globalized attention calculation on the temporal trends of multiple nodes.


\end{itemize}

\begin{figure*}[htbp]
\subfigure[]{
\includegraphics[width=1\columnwidth]{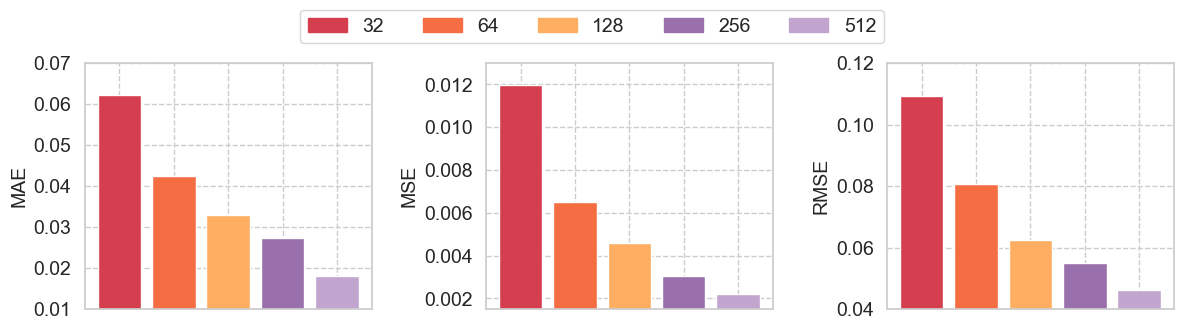}
\label{dmodel}}
\subfigure[]{
\includegraphics[width=1\columnwidth]{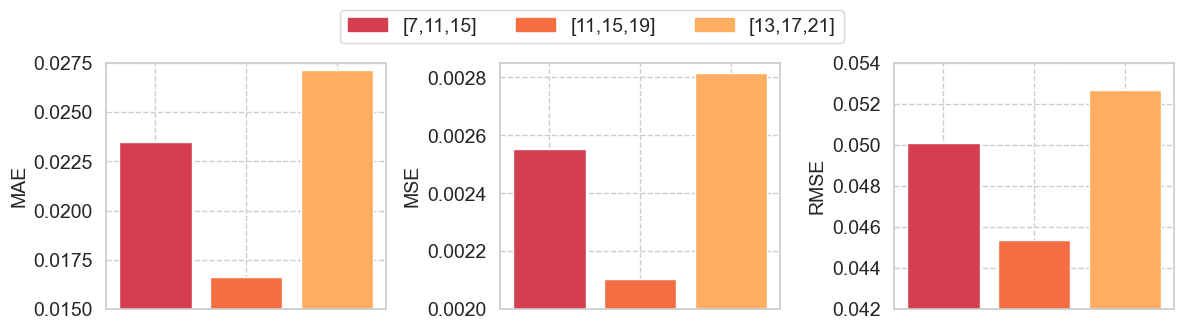}
\label{decomp}}
\subfigure[]{
\includegraphics[width=1\columnwidth]{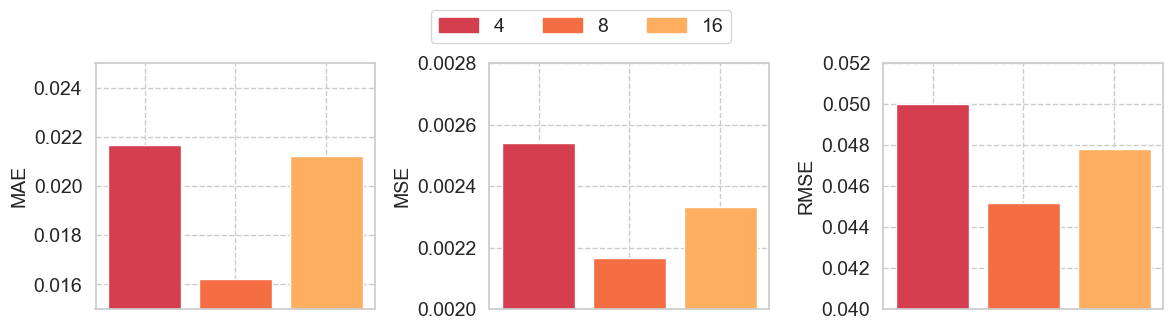}
\label{n_h}}
\subfigure[]{
\includegraphics[width=1\columnwidth]{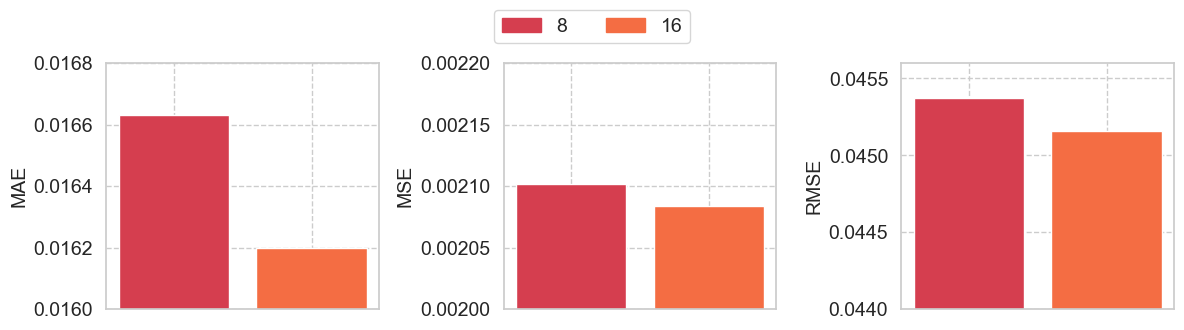}
\label{factor}}
\subfigure[]{
\includegraphics[width=1\columnwidth]{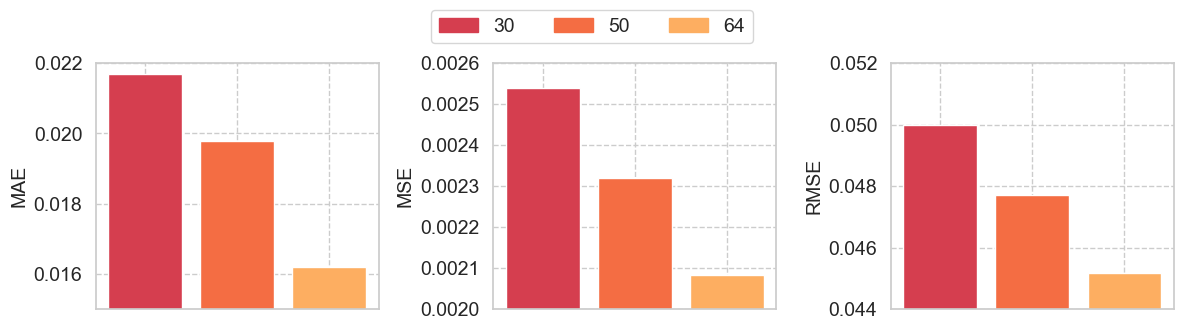}
\label{pca_m}}
\subfigure[]{
\includegraphics[width=1\columnwidth]{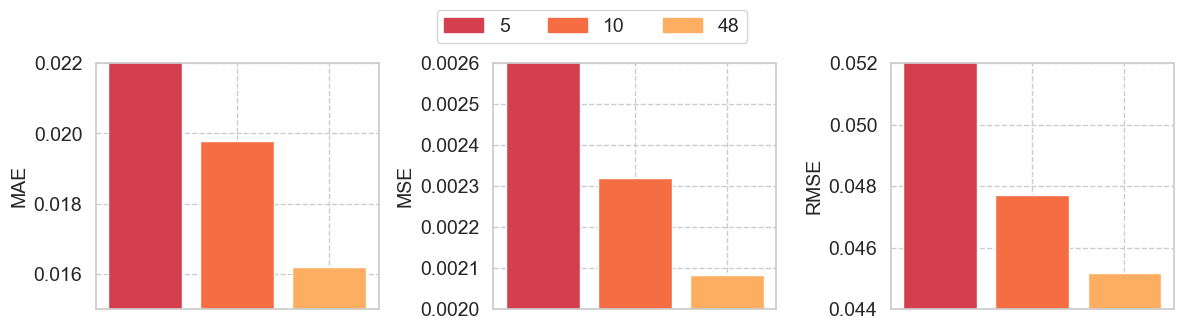}
\label{pca_k}}
\caption{Impact of hyperparameters on model performance. (a) evaluation based on different dimension of model. (b) evaluation based on the choice of kernel length for sequence decomposition. (c) evaluation based on the number of attention heads. (d) evaluation based on the number of attention heads. (e) evaluation based on the dimension of random feature mapping in TMM-PCA. (f) evaluation based on the number of sampling for gumble-softmax in TMM-PCA. (32 prediction steps)}
\Description[]{}
\end{figure*}

As the Table \ref{ablation} shows, the whole model is superior to other variants. Based on the ablation experimental results, we can obtain following conclusions: (1) The system's topological deployment information can aid in predictions, with close relationships existing both among instances on the same host and between hosts and instances. Removing the IMM leads to a significant increase in prediction errors. This is because a noticeable change in the state of the host or one of its instances can disrupt the operation of other instances. (2) Dynamic adjacency matrices significantly impact system state predictions. Removing the SMM also increases errors. This is because variations in TCP connection events at each time step can alter the dependencies among instances, thereby affecting their states. Removing this module would omit the transmission of spatial information within single time step. (3) Removing TMM-TB would increase prediction errors. The primary reason is that the variables have hidden trends within the time series that follow certain patterns, which are not solely influenced by spatial relationships. This highlights the necessity to account for time-intrinsic fluctuation in addition to spatial interactions. (4) Calculating the impact of cascading effects from a global perspective aids in system state prediction. After the removal of the TMM-PCA, an increase in prediction errors was observed. Computing attention from a global spatio-temporal perspective effectively captures the state transmission between multi-hop nodes and over long time intervals. This approach highlights the necessity of incorporating a global analysis to understand and predict the dynamics across the network of interconnected services more accurately.

\subsection{Hyperparameter Studies}
Changes in the hyperparameters of the STMformer typically affect model's predictive performance. To validate the impact of hyperparameters on the STMformer, we will analyze from five perspectives: the dimension of the model, the choice of kernel length for sequence decomposition, the number of attention heads, the ProbSparseAttention factor, and the hyperparameters of the TB-PCA module.

As the Fig.~\ref{dmodel} shows, changes in model parameters significantly affect the performance of the STMformer. As the dimensions of model parameters increase, the performance of the model improves. This indicates that a higher-dimensional vector space enhances the model's learning capabilities. Node states, after being processed through various spatio-temporal relationship calculations, implicitly contain elements of physical spatial deployment, spatial logical relationships, trends in time series, and global spatio-temporal relationships. Enhancing model parameters helps nodes learn these complex features. Fig.~\ref{decomp} illustrates that the model is influenced by the choice of kernel for sequence decomposition. The optimal selection of kernel length depends on the pattern fluctuations of the time series and expert experience. Choosing the appropriate set of kernel lengths can more accurately obtain trends, allowing the model to focus on learning the remaining parts of the series.

The Fig.~\ref{n_h} and Fig.~\ref{factor} display experiments related to attention-related hyperparameters. The choice of the number of attention heads, if too large or too small, affects the model's performance, as it influences the model's ability to parallelize learning across different aspects of the data, potentially leading to either redundant learning or insufficient feature capture. Regarding the selection of the ProbSparseAttention factor, it is evident that its impact on model performance is not significant. This is because, in the context of instance states, only a small portion of the variables are affected. ProbSparseAttention consistently selects the variables with the highest associations, ensuring that the most critical features are always considered.

The Fig.~\ref{pca_m} and Fig.~\ref{pca_k} indicate that higher dimensions of random feature mapping and more sampling times of Gumbel-Softmax correlate with improved model performance. Increasing the dimensionality of mapping means that spatio-temporal information can be better represented in another vector space. Meanwhile, increasing the number of sampling times means that the operation's frequency is higher, ensuring more accurate approximation of the softmax distribution. Although these two hyperparameters increase computation time, such a trade-off is acceptable compared to the benefits of achieving better model performance.




\section{Conclusion and Future Work}

For predicting the state of microservices systems, we developed the STMformer model, which leverages dynamically derived adjacency matrices from network events and deployment structures to capture spatio-temporal relationships characteristic of microservices systems. We also design PatchCrossAttention module to analyze cascading effects from a global spatio-temporal perspective. Our experiments show that STMformer outperforms benchmark methods, confirming the effectiveness of its components and the influence of hyperparameters on its performance. Our future work will focus on reducing computational complexity, automating kernel generation for sequence decomposition, and integrating multimodal data for more robust microservices system representations.

\begin{acks}
This work is partially supported by the \grantsponsor{52071312}{National Natural Science Foundation of China}{} under Grant No.\grantnum{52071312}{52071312} and \grantsponsor{2022YFB3103402}{National Key Research and Development Program}{} under Grant No.\grantnum{2022YFB3103402}{E250411105}.
\end{acks}
%
%

\clearpage
\balance
\bibliographystyle{ACM-Reference-Format}
\bibliography{reference}

\end{document}